\documentstyle[emulateapj,psfig]{article}
\def\simlt{\lower.5ex\hbox{$\; \buildrel < \over \sim \;$}}

\begin{document}

\title{QSO 2359$-$1241:  A Bright, Highly Polarized, Radio-Moderate, 
Reddened, Low-Ionization Broad Absorption Line Quasar}
\author{M. S. Brotherton\altaffilmark{1,2}, Nahum Arav\altaffilmark{3}, R. H. Becker\altaffilmark{1,3}, Hien D. Tran\altaffilmark{4}, Michael D. Gregg\altaffilmark{1,3}, R. L. White\altaffilmark{5}, S. A. Laurent-Muehleisen\altaffilmark{3},\& Warren Hack\altaffilmark{5}}

\altaffiltext{1}{Institute of Geophysics and Planetary Physics, Lawrence Livermore National Laboratory, 7000 East Avenue, P.O. Box 808, L413, Livermore, CA 94550}
\altaffiltext{2}{Kitt Peak National Observatory, National Optical Astronomy Observatories\footnote{The National Optical Astronomy Observatories are operated by the Association of Universities for Research in Astronomy, Inc., under cooperative agreement with the National Science Foundation.},  950 North Cherry Avenue, P. O. Box 26732, Tuscon, AZ 85726}
\altaffiltext{3}{Department of Physics, University of California, Davis, CA 95616}
\altaffiltext{4}{Department of Physics and Astronomy, Johns Hopkins University, Baltimore, MD 21218}
\altaffiltext{5}{Space Telescope Science Institute, 3700 San Martin Drive, Baltimore, MD 21218}
\begin{abstract}

We report the discovery of a bright quasar ($E=15.8, z=0.868$) 
associated with the flat spectrum radio source NVSS J235953$-$124148.  
This quasar we designate QSO 2359$-$1241 possesses a rare combination of 
extreme properties that make it of special interest.  These properties
include:  intrinsic high-velocity outflow seen in absorption for both
high and low-ionization species, high optical polarization ($\sim5\%$),
significant radio emission, and dust reddening.  The dereddened
absolute magnitude of QSO 2359$-$1241 places it among the three
most optically luminous quasars known at $z<1$.
High-resolution spectroscopy and a detailed analysis of the
optical/ultraviolet absorption features will be given in a companion paper
(Arav et al 2000).  

\end{abstract}

\section{Introduction}

Rare objects with extreme properties sometimes prove exceptionally
useful as case studies for unraveling the physics underlying those 
extreme properties.  This maxim is amplified when these objects are 
bright and easily observed.

Quasars are rare, displaying the extremes of high 
luminosity that galactic nuclei are capable of achieving.
This makes them potentially useful as cosmological probes, 
especially if the supermassive black holes of quasars trace the earliest mass 
concentrations (e.g., Loeb 1993) or if the ignition of quasars traces
galaxy mergers (e.g., Sanders et al. 1988; Hernquist 1989; Mihos \& Hernquist
1996) that are the signatory events of heirarchical cosmogonies 
(e.g., Turner 1999).

Among optically selected quasars, some $\sim$10\% are radio-loud
(log R* $>$ 1, where R* is the $K$-corrected ratio of radio-to-optical power;
Sramek \& Weedman 1980; Stocke et al. 1992).
Only one optically selected quasar in a hundred shows
blueshifted broad absorption lines from Mg II characteristic of
a low-ionization outflow.  Of these ``LoBAL'' quasars, only a small
fraction shows high optical polarization greater than 3\%
(e.g., Schmidt \& Hines 1999; Ogle et al. 1999; Hutsemekers, Lamy, 
\& Remy 1998); just what fraction is uncertain due to the small
size of samples so far studied and heterogenous selection.
Furthermore, most quasars are blue and most easily selected using
ultraviolet-excess techniques; selection methods without optical
color biases shows only a small population of bright red quasars
(e.g., Gregg et al. 1996).

While the properties of optical polarization, red color, and
low-ionization absorption are rare, they often appear together.
Moreover, quasars with these properties universally have strong
optical Fe II emission and negligible emission from narrow forbidden
lines, notably [O III] $\lambda$5007 (see Voit et al. 1993 and 
Turnshek 1996).  These objects form an extreme end of quasar 
properties, an extreme of ``eigenvector 1'' of Boroson \& Green (1992),
and include objects such as I Zw 1, IRAS 07598+6508, and narrow line Seyfert 1
galaxies.  Voit et al. (1993) and Boroson \& Meyers
(1992) proposed that these objects were enshrouded quasars with large
covering factors, perhaps young emerging objects.  Laor et al. (1997)
favored the idea that these objects represent high accretion rate
objects.  Whatever the cause of the extreme spectral properties, it
is apparently a fundamental parameter governing the appearance of quasars. 

The most extreme case to date of a red, absorbed, polarized quasar is
FIRST J155633.8+351758 (Becker et al. 1997).  Its peak polarization
of 13\% (Brotherton et al. 1997) is among the largest for a broad 
absorption line quasar, its absorption is among the heaviest, it
is among the reddest of quasars ($B-K = 6.57$; Hall et al. 1998),
and it too has strong optical Fe II and weak narrow line emission
(Najita et al. 2000).  Becker et al. (1997) emphasized the then uniquely high 
radio luminosity of FIRST J155633.8+351758 which places it among radio-loud
quasars, although after correcting for apparent reddening the log R* measure
categorizes this object as radio moderate.  The moderate reddening
and redshift ($z=1.5$) conspire to make this quasar faint in optical
bands ($V \sim 19$).  This object has been proposed as the tip of an
iceberg of red, dusty absorbed objects, akin to the earlier suggestion
by Webster et al. (1995) of a large reddened class of quasars; 
Chartas (2000) notes that approximately 35\% of radio-quiet gravitational
lenses contain BAL features and suggests that flux-limited optical surveys
miss a large fraction of BAL quasars.
The faintness and heavy blending of the numerous strong absorption 
lines has made FIRST J155633.8+351758 difficult to study in many
respects.

We have discovered another member of the red, polarized LoBAL quasar
class also bright in the radio.  Moreover this new object is quite
optically bright, which permits a variety of detailed studies.  Below we report
the discovery and follow-up observations in the radio and ultraviolet bands,
as well as optical spectropolarimetry. 
Detailed analysis of the ultraviolet absorption
lines and high-resolution optical spectroscopy will be presented in
a companion paper (Arav et al. 2000; Paper II).  
We adopt $H_0 = 50$ km s$^{-1}$ Mpc$^{-1}$, $q_0$ = 0, $\Lambda=0$
throughout this paper.

\section{Discovery and Follow-up Observations}

We have been engaging in several programs of identifying quasars associated
with radio sources in large, deep surveys such as FIRST (Faint Images of the
Radio Sky at Twenty centimeters; Becker et al. 1995) and the NVSS
(NRAO\footnote{The National Radio Astronomy Observatory is a facility of
the National Science Foundation operated under cooperative agreement by
Associated Universities, Inc.} VLA Sky Survey; Condon et al. 1998).
By matching radio sources to optically bright stellar sources (Gregg et al. 
1996; Brotherton et al. 1998; White et al. 2000; McMahon et al. 2000), 
quasars can be efficiently located.  Such a procedure has proven successful
at selecting radio-loud BAL quasars and large numbers of radio-intermediate 
quasars apparently overlooked using purely optical techniques 
(White et al. 2000).

We targeted the bright stellar source ($O$= 16.7, $E$=15.8) at
$\alpha =$ 23$^h$59$^m$53$\fs$6, $\delta = -$12$\arcdeg$ 41$'$49$\farcs$0 
(J2000),
coincident with NVSS J235953$-$124148 (unresolved radio source with
flux density of 39.8$\pm$1.3 mJy at 20 cm) in a pilot program on 
9 February 1995 at Keck Observatory using the LRIS spectrograph 
(Oke et al. 1995).  The spectrum (Figure 1, $\lambda_{observed} > 4000 \AA$) 
reveals a $z=0.868$ quasar we designate as QSO 2359$-$1241, which
additionally displays two deep absorption troughs blueward of the 
Mg II $\lambda$2800
emission line. The optical and radio fluxes yield a radio-loudness
of log R* = 1.66 (using the same slopes for K-corrections as do
Stocke et al. 1992), making QSO 2359$-$1241 formally radio-loud.  
The Galactic extinction toward this object is a minimal $E(B-V)=0.029$\ 
(Schlegel, Finkbeiner, \& Davis 1998) 
and does not significantly affect these numbers. 

In the remainder of this section we describe our follow-up observations of
this quasar, including radio imaging, optical spectropolarimetry, and
ultraviolet spectroscopy\footnote{Based on observations with the NASA/ESA 
Hubble Space Telescope, obtained at the Space Telescope Science Institute, 
which is operated by the Association of Universities for Research in Astronomy,
Inc. under NASA contract No. NAS5-26555.}.
The analysis of the high-resolution spectroscopy, and the
detailed comparison of optical and ultraviolet absorption features, is 
performed by Arav et al. (2000).

\subsection{Radio}


We obtained follow-up radio observations with the VLA in A-array.
The 20 cm flux density peak was 37.3 mJy (17 March 1998), 
and the 3.6 cm flux density peak was 19.8 mJy (16 March 1998).
The source was unresolved at 3.6 cm (beamsize 0.40 x 0.23 arcsec) and 
marginally resolved at 20 cm (beamsize 1.92 x 1.33 arcsec).
This yields a radio spectral index of $\alpha = -0.36$
(where $\alpha$ is defined by S$_{\rm \nu}\propto\nu^{\alpha}$). 
Bandwidth was the standard 50 MHz, and the source's absolute flux was 
determined using 3C 48.  The position on the 3.6 cm (better resolution) map is 
$\alpha =$ 23$^h$59$^m$53$\fs$627, $\delta = -$12$\arcdeg$41$'$47$\farcs$92
(J2000), in agreement with the optical position.

\subsection{UV Spectroscopy}

We obtained an ultraviolet spectrum using the Hubble Space Telescope's
f/96 Faint Object Camera (FOC) with an objective prism on 1999 June 30
with an exposure time of 956s.  The objective prism image was taken with
the PRISM2 near-UV objective prism (NUVOP) and provided low-resolution
spectroscopy.  The observations were flat-fielded and geometrically
corrected using the standard FOC pipeline calibrations described in detail
in Nota et al. (1996).  The dispersed images taken with the objective
prism are aligned with the visible wavelengths on the lower end and the UV
wavelengths at the upper end.  The NUVOP image has a resolution of 40 
\AA\ pixel$^{-1}$ at 5000 \AA\ and 0.5 \AA\ pixel$^{-1}$ at 1700 \AA.  
A 7 pixel wide region
centered on the peak of the spectrum was extracted and summed to create
the raw spectrum.  This was then converted from counts per pixel to flux
per Angstrom by applying the latest dispersion relation for the NUVOP and
encircled energy percentages for the extracted spectrum as given in the
FOV Instrument Handbook (Nota et al.  1996).  The photometry and
wavelength calibrations were verified by matching common spectral features
and fluxes with the ground-based spectra and APM photometry.
The portion of Figure 1 with $\lambda_{observed} < 4000 \AA$\
is the FOC spectrum.

\subsection{Spectropolarimetry}


We observed QSO 2359$-$1241 on 21 July 1999 (UT) with the Low Resolution
Imaging Spectrometer (Oke et al. 1995) in spectropolarimetry mode
(Good\-rich, Cohen, \& Putney 1995; Cohen et al. 1997) on the 10 meter Keck 
II telescope.
We used a 300 line mm$^{-1}$ grating blazed at 5000 \AA,
that, with the 1 $\arcsec$ slit (at the parallactic angle),
gave an effective resolution of 10 \AA\ (FWHM of lamp lines);
the dispersion was 2.5 \AA\ pixel$^{-1}$.
The seeing was $\sim1.2^{\prime\prime}$.
The observation was broken into four 10 minute exposures, one for each
waveplate position (0$\arcdeg$, 45$\arcdeg$, 22.5$\arcdeg$, 67.5$\arcdeg$).
The red end of the spectrum ($\lambda_{obs} >$ 7400\AA) is weakly
contaminated by second-order light at a level of $\simlt$ 5\%
(see e.g., Ogle et al. 1999).

We reduced our data to one-dimensional spectra using standard techniques within
the IRAF NOAO package. The rms uncertainties in the dispersion solution were
0.2\AA, and we used sky lines to ensure that our zero point was accurate to
0.1\AA.  Wavelengths are air wavelengths.  We followed standard procedures
(Miller, Robinson, \& Goodrich 1988; Cohen et al. 1997) for calculating
Stokes parameters and uncertainties.

Figure 2 plots our results, showing the total flux spectrum, linear
polarization, position angle, and polarized flux.  For clarity we have
omitted error bars, but the signal-to-noise ratio is high even without
binning.  However for a detailed look at the polarization across the Mg II
emission and absorption features, Figure 3 plots a blow-up of this region
with 10 \AA\ binning (done in Stokes $q$\ and $u$\ prior to calculating $p$)
and 1 $\sigma$\ error bars.

\section{Results \& Analysis}

The first result to focus on is that this quasar does indeed possess 
broad absorption lines.  While the high polarization, red color, and
deep Mg II absorption troughs are suggestive of QSO 2359$-$1241 being
a low-ionization BAL quasar, the absorption troughs themselves in the 
observed-frame optical are insufficient to make this classification.  
The ultraviolet spectrum shows BAL troughs covering more than 8,000 km s$^{-1}$.
Continuum placement is difficult, but we estimate that the C IV trough
has a BALnicity (Weymann et al. 1991) of $\sim$2300 km s$^{-1}$.  
Clearly the outflow displays a stratification in ionization and velocity, 
which is characterized and discussed in detail in Paper II.

The spectropolarimetric properties are somewhat unusual but not atypical of BAL 
quasars.  QSO 2359$-$1241 is linearly polarized at $\sim$4\% and a position 
angle of $\sim$140$\arcdeg$ at 8600 \AA\ observed frame, rising to 
$\sim$6\% and $\sim$150$\arcdeg$ at 4000 \AA\ observed frame.
The broad emission line (i.e. Mg II) appears unpolarized, and the 
polarization does not change very significantly across the absorption troughs.

The total light spectrum has a redder spectrum than that of the typical
blue quasar.  Additionally the manner in which the linear polarization 
rises toward shorter wavelengths is suggestive of a diminishing amount of
dilution, as from a reddened direct light spectrum (as in 3CR 68.1,
Brotherton et al. 1998).  Using the color indexes defined by
Yamamoto \& Vansevicius (1999), $F_{1750}/F_{2000} \approx 0.9$ and
$F_{2100}/F_{3000} \approx 1$, and their diagnostic diagrams, 
QSO 2359$-$1241 would appear to be 
reddened by $E(B-V) \approx 0.25$ using a Small Magellanic Cloud (SMC) 
extinction law as compared to the composite quasar spectrum of 
Francis et al. (1991).  We have explored the relative reddening of
QSO 2359$-$1241 as compared to a composite formed from quasars in
the FIRST Bright Quasar Survey (Brotherton et al. 2000).  Figure 4 
shows that dereddening QSO 2359$-$1241 with an SMC-type extinction
law (Prevot et al. 1984) and $A_V = 0.5$ produces a good match.
The match at the shortest wavelengths is complicated by the strong absorption 
and an increasing fraction of scattered light suggested by
the rising polarization.

The apparent reddening of QSO 2359$-$1241 has implications for its
luminosity and radio-loudness.  QSO 2359$-$1241 is bright and at moderate
redshift, already indicating it is a very luminous quasar with 
M$_B$ = $-$27.9.  Dereddening boosts its luminosity to M$_B$ = $-$28.7.
Figure 5 compares these numbers to those of the more than 11000 quasars
in the Veron-Cetty \& Veron (1998) catalog, revealing that a dereddened
QSO 2359$-$1241 would be the third most luminous quasar at $z<1$.

The value of log R* also drops with an increased intrinsic optical
luminosity, to log R* = 1.0 -- leaving it straddling the commonly used
radio-quiet/radio-loud division.  The 5 GHz rest-frame luminosity 
is L$_R = 10^{33.7}$ erg s$^{-1}$ Hz$^{-1}$, which is still more than
an order of magnitude above the radio-quiet/radio-loud division of 
L$_R = 10^{32.5}$ erg s$^{-1}$ Hz$^{-1}$ proposed by Miller, Rawlings, 
\& Saunders (1993).  The large radio emission certainly comes from the quasar,
but it would appear that QSO 2359$-$1241 is overly optically luminous.

\section{Discussion}

QSO 2359$-$1241 is certainly a rarity among rarities.
It is a radio-loud BAL quasar (although radio moderate is probably
a more appropriate description as the evidence for a radio-quiet/radio-loud
dichotomy has lessened; White et al. 2000).  It is extremely luminous
for a low redshift quasar, reminiscent of the BAL quasar 
APM 08279+5255 -- probably the most luminous object in the known universe
(e.g., Irwin et al. 1998) even after correction for lensing amplification.  
It is also reminiscent of the first
discovered radio-loud BAL quasar, FIRST 1556+3517 (Becker et al. 1997),
which also shows low-ionization absorption, colors suggestive of dust
reddening (Clavel 1998; Hall et al. 1998; Najita et al. 2000), 
and very high optical polarization (Brotherton et al. 1997).

The polarization mechanism is very likely a scattering process,
as has been proposed and discussed previously for other
BAL quasars (Ogle et al. 1999; Schmidt \& Hines 1999; Brotherton et al. 1997; 
Hines \& Wills 1995;
Cohen et al. 1995; Goodrich \& Miller 1995; Glenn, Schmidt, \& Foltz 1994).  
The contribution from Galactic interstellar 
polarization is negligibly small here given the very low E(B-V) along 
this line of sight.  While often in BAL quasars the polarization rises
in absorption troughs (presumably because the scattered line of sight
is less absorbed than the direct line of sight), it is not the 
case in QSO 2359$-$1241; higher resolution may be required to 
see such a rise from the rather narrow Mg II absorption features
(see Arav et al. 2000).
The fact that the broad emission lines are not polarized, which is
also the case for the majority of BAL quasars, suggests that the 
scattering medium is coincident with or smaller than the broad line region.
The rise in the polarization toward smaller wavelengths is also commonly
seen in BAL quasars, as well as red non-OVV quasars (e.g., 3CR 68.1,
Brotherton et al. 1998).  A strong hypothesis is that the scattered
line of sight is less reddened than that to the polarization-diluting 
direct continuum.  Finally, the change in position angle with 
wavelength is uncommon but previously noted in BAL quasars
(e.g., Ogle et al. 1999) and suggestive of a complex geometry in which
there are multiple scattered lines of sight with different amounts
of reddening.

Arav et al. (2000) show that the Mg II absorption only partially covers
the continuum source, and that absorption features of different 
ionization show different velocity structures.  As has been pointed
out (e.g., Voit et al. 1993), lower ionization species in BAL outflows
appear at lower velocities.  The flows and geometries of these quasars
are complex.

QSO 2359$-$1241 has a compact radio morphology like the vast majority of
other radio-selected BAL quasars (Becker et al. 2000), and like
FIRST 1556+3517 it has a flat spectrum ($\alpha < 0.5$).  Both of these
properties make it likely that the geometry of QSO 2359$-$1241 is
not ``edge-on'' to the system axis, which has been a popular model
for polarized BAL quasars and appears to be a very successful idea for
explaining polarized Seyfert 2 galaxies that habor hidden Seyfert 1 nuclei.

Low-ionization BAL quasars like QSO 2359$-$1241 so far observed at
rest-frame optical wavelengths all have very weak or negligible
[O III] $\lambda$5007 and other lines from extended narrow line regions
and have strong optical Fe II emission.  QSO 2359$-$1241 also
shows strong optical Fe II emission (the multiplet near 4570 \AA\ rest frame).
These properties are also found in Narrow Line Seyfert 1 galaxies
and characterize one extreme of Boroson \& Green's ``eigenvector 1'' 
which accounts for the most variance in the optical spectra of 
low-redshift quasars.  Boroson \& Green (1992) also argued that the 
Eddington fraction was the important parameter underlying the
[O III] $\lambda$5007 -- Fe II anti-correlation dominated eigenvector 1.  
They surmised that optical Fe II emission was
dependent on the covering fraction of the BLR, and that more BLR clouds
(and hence higher accretion rate) would obscure the more distant NLR.
Thus the covering fraction increases from the radio-loud strong [O III]
$\lambda$5007, weak Fe II quasars to the radio-quiet weak
[O III] $\lambda$5007, strong Fe II quasars.  They also noted that PG
1700+518, a BAL quasar, is found at the high
covering fraction end.  Laor et al. (1997) also argued in favor of 
this explanation after characterizing how the X-ray slope becomes 
softer (with an increasing excess) for quasars with small [O III]/Fe II.

X-rays are exceedingly weak from BAL quasars (Green \& Mathur et al. 1996).
Recently Gallagher et al. (1999) reviewed the X-ray properties of eight 
BAL quasars targeted by {\em ASCA}: only half were detected despite
the hard energy sensitivity, and assuming normal quasar X-ray properties
these non-detections imply columns with $N_{\rm H} \geq 5\times10^{23} cm^{-2}$.
We have been allocated {\em Chandra} time to observe QSO 2359$-$1241 and
will attempt to determine the intrinsic X-ray properties and those of the
absorber likely to be present.  

We speculate that low-ionization BAL quasars like QSO 2359$-$1241,
and other quasars at this extreme of Boroson \& Green's eigenvector 1,
represent young or recently refueled quasars with high accretion rates
and high covering fractions (either from infalling fuel from a triggering
event or outflowing debris from the engine running near the Eddington 
limit).  If quasars dim as they age and fuel runs out, 
we may expect a high fraction of the most luminous quasars to be 
BAL quasars.

\section{Summary}

We have reported the discovery of the bright radio-moderate quasar 
QSO 2359$-$1241.  This quasar is highly polarized and reddened, and 
possesses an intrinsic high-velocity outflow with a complex ionization/velocity
structure (investigated in detail by Arav et al. 2000).

\acknowledgments

We thank the Arjun Dey and Dean Hines for their assistance.
The National Radio Astronomy Observatory is a facility of the 
National Science Foundation operated under cooperative agreement by Associated 
Universities, Inc.  
The W. M. Keck Observatory is a scientific partnership 
between the University of California and the California Institute of Technology,
made possible by the generous gift of the W. M. Keck Foundation.
Support for this work was provided by NASA through grant number GO-06350 
from the 
Space Telescope Science Institute, which is operated by AURA, Inc., under NASA 
contract NAS5-26555.  We acknowledge support from the National Science 
Foundation under grant AST 98-02791.
This work has been performed under the auspices of the U.S. Department of Energy
by Lawrence Livermore National Laboratory under Contract W-7405-ENG-48.




\newpage
\psfig{file=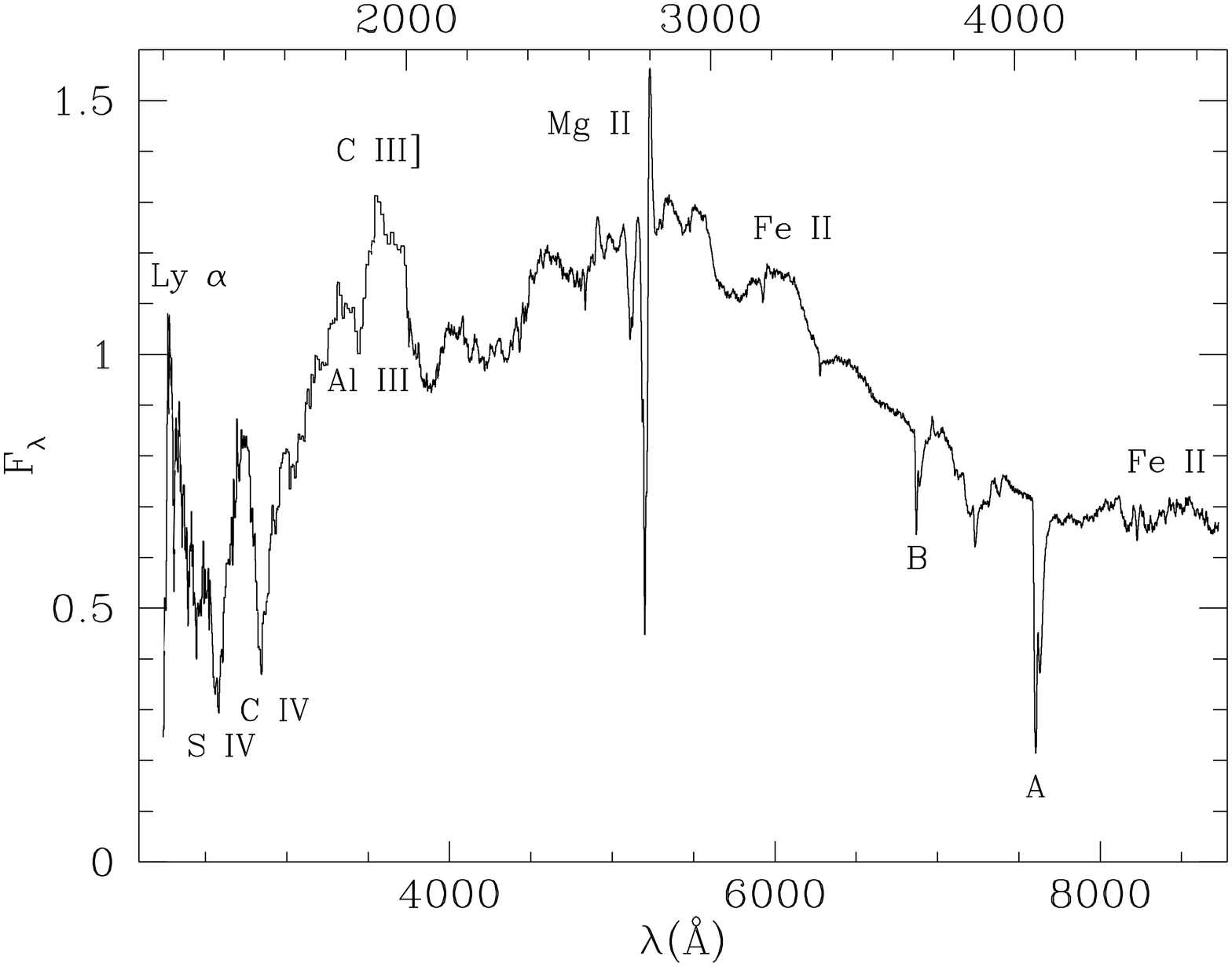,angle=-90,height=14cm}
\figcaption{
The ultraviolet portion of the {\em Hubble Space Telescope} FOC spectrum
combined with the optical Keck discovery spectrum.
The flux in units of ergs cm$^{-2}$ s$^{-1}$ \AA$^{-1}$ have been
multiplied by $10^{15}$.  The wavelengths for the observed frame (bottom axis)
and rest frame (top axis) are shown in \AA.
}


\newpage
\psfig{file=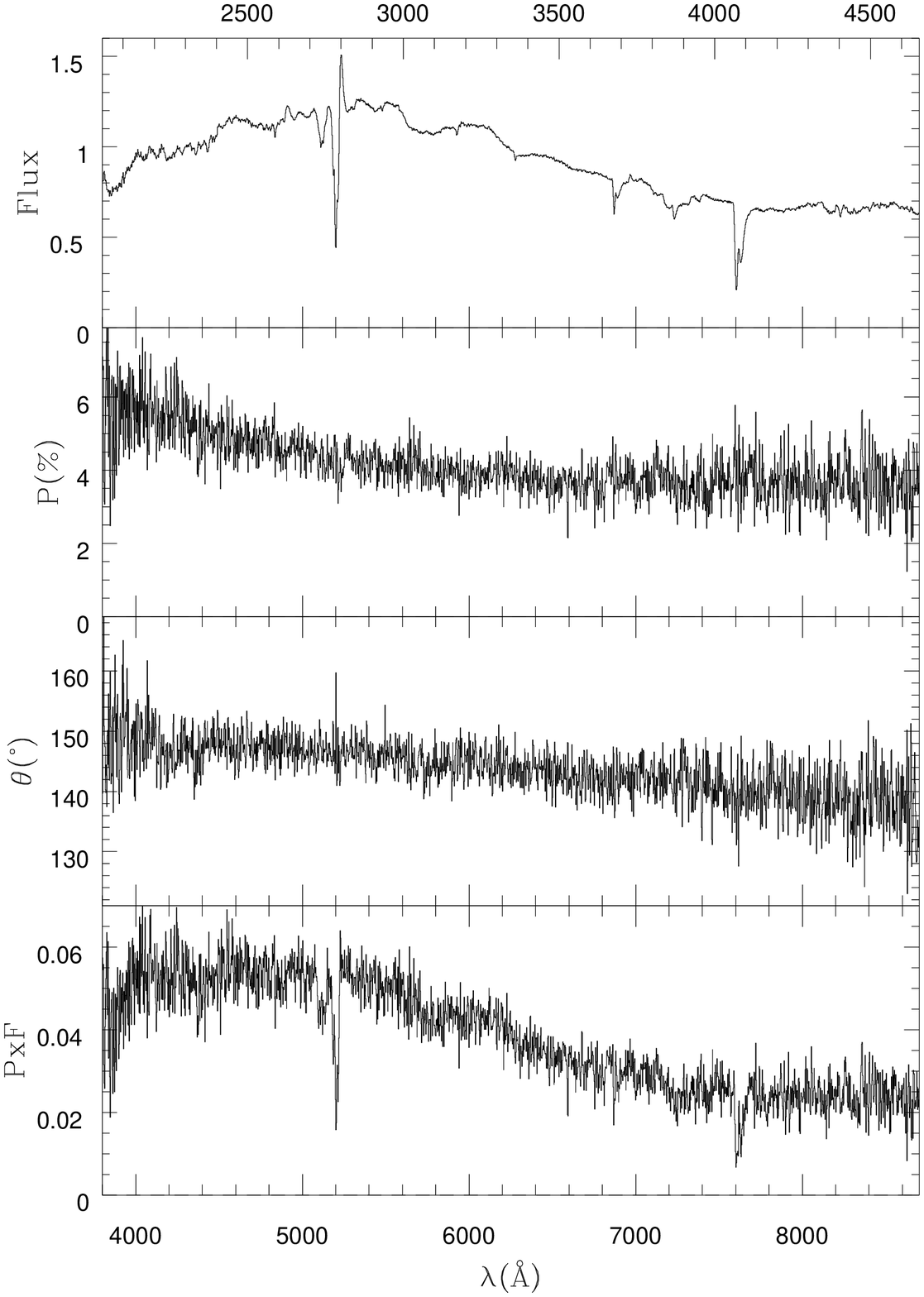,height=22cm}
\figcaption{
Spectropolarimetric results.
The top panel is the total flux spectrum.
The second panel from the top shows the percentage polarization.
The third panel is the polarization position angle in degrees.
The bottom panel shows the polarized flux, the product of the 
fractional polarization and total light spectrum.
The fluxes (in units of ergs cm$^{-2}$ s$^{-1}$ \AA$^{-1}$) have been
multiplied by $10^{15}$.  The wavelengths for the observed frame (bottom axis)
and rest frame (top axis) are shown in \AA.
}

\newpage
\psfig{file=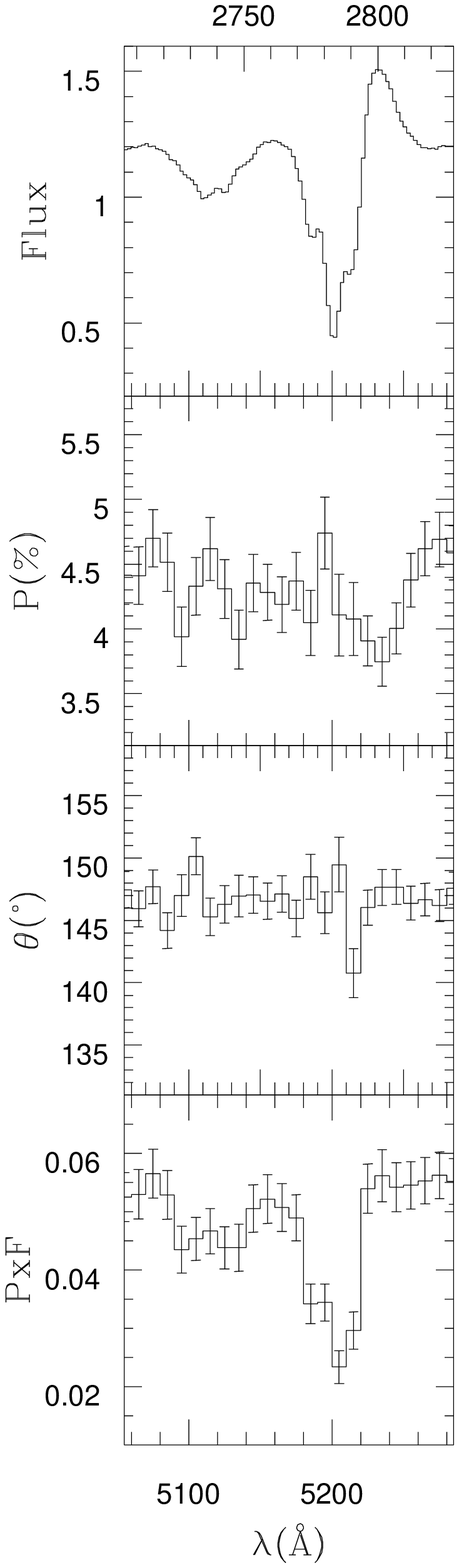,height=22cm}
\figcaption{
A closer look at the spectropolarimetry near the Mg II region, 
10 \AA\ binning (resolution element), 1 $\sigma$ error bars.
The panels, units, and labels, are the same as Figure 2.
}

\newpage
\psfig{file=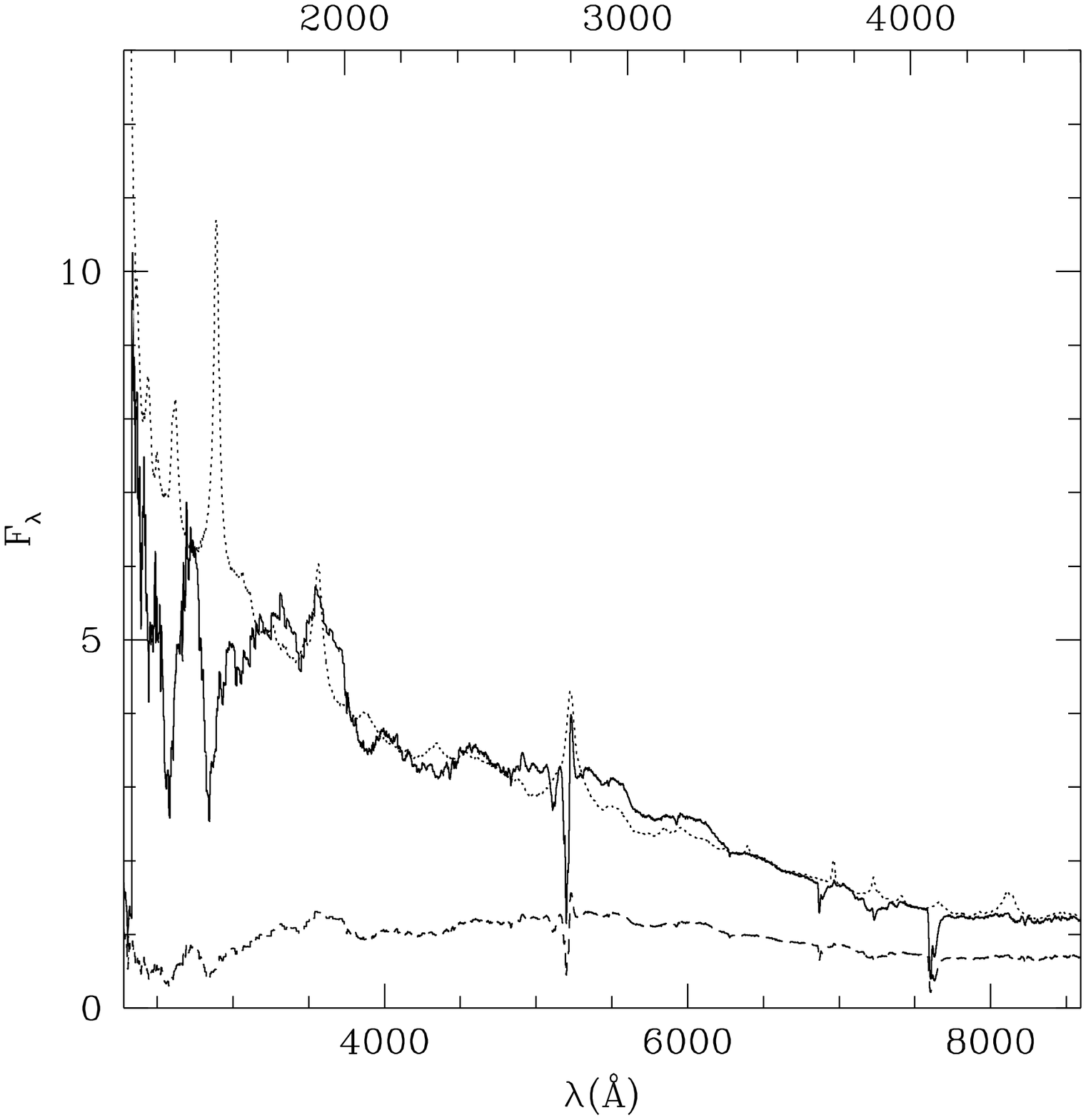,height=18cm}
\figcaption{
The bottom spectrum (dashed line) is the observed spectrum (as in Figure 1).  
The solid line above is this spectrum after dereddening 
by $A_V=0.5$ with an SMC-extinction law.  The dotted line is the
FIRST Bright Quasar Survey composite spectrum of Brotherton et al. 2000.
The fluxes (in units of ergs cm$^{-2}$ s$^{-1}$ \AA$^{-1}$) have been 
multiplied by $10^{15}$.  The wavelengths for the observed frame (bottom axis)
and rest frame (top axis) are shown in \AA.
}

\newpage
\psfig{file=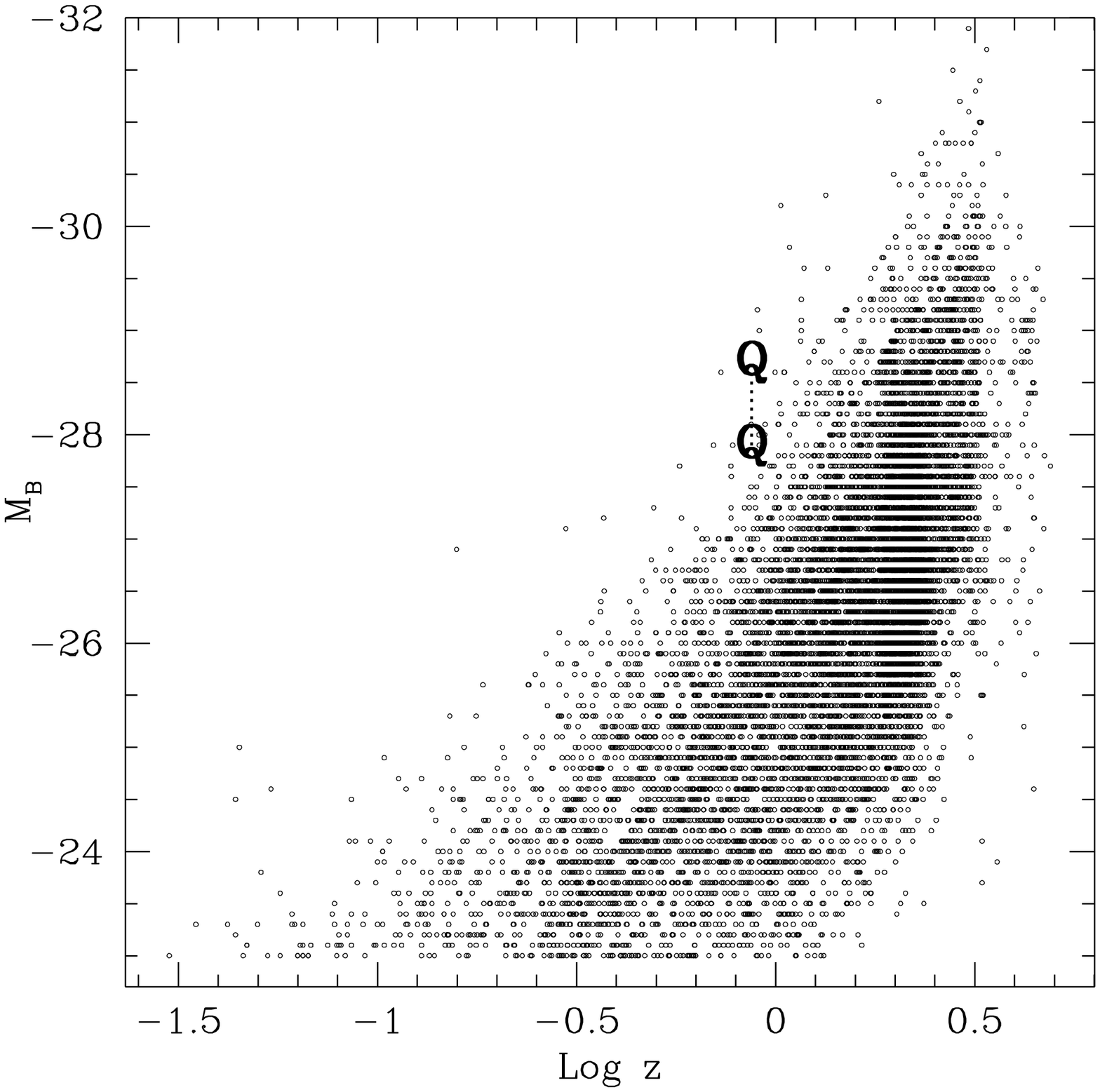,height=18cm}
\figcaption{
Comparison of the as observed (lower) and dereddened (upper) absolute 
magnitudes of QSO 2359$-$1241 (labelled with "Q") to the more than
11,000 quasars in the Veron-Cetty \& Veron (1998) catalog.  Dereddened,
QSO 2359$-$1241 would appear to be the third most optically luminous
$z<1$\ quasar.
}

\end{document}